# The energies and ANCs for $^5$Li resonances deduced from experimental *p-α* scattering phase shifts using the effective-range and Δ methods

Yu.V. Orlov[*]

*Skobeltsyn Nuclear Physics Institute, Lomonosov Moscow State University, Russia*

(Dated: August 26, 2020)

**Abstract**

Recently a new Δ method for deducing the energy and asymptotic normalization coefficient (ANC) from phase-shift data has been formulated and applied to resonance states. This differs from the conventional effective-range function (ERF) method by fitting only the nuclear part of the ERF. It also differs from the method which was proposed for bound states by Ramírez Suárez and Sparenberg (2017) which also named the Δ method where a pole condition defines by the Eq. $\Delta_l$=0 ($\Delta_l$ is the function in the ERF determined only by the scattering phase shift). Here the standard pole condition, including the Coulomb part in the relate equation, is used for a resonant state. It has been shown that the ERF method does not work for large-charge colliding nuclei. Moreover, even for lower charges it is not clear that the results of the ERF method are accurately enough. The Coulomb part forms a background, which smooths an ERF energy dependence. Therefore, one needs to find when the ERF method becomes inaccurate and this requires recalculating some published results by the Δ method. This project has already been started in a recent paper for resonances in the *α-α* scattering. Here this method is applied using the $\Delta_l$-function fittings to the experimental *p*-$^4$He scattering phase-shift data in the $P_{3/2}$ and $P_{1/2}$ resonance states. The calculation results are compared with those obtained earlier by the ERF method. The main changes concern resonance energy and width.

Keywords : $^5$Li states *l*=1, J=1/2, 3/2; *p*-$^4$He elastic scattering phase-shift input data; Low-energy resonance; Asymptotic normalization coefficient (ANC); Effective-range and Δ methods calculation comparison; Nuclear astrophysics applications; Radiative capture cross sections at low energies

[*] orlov@srd.sinp.msu.ru

## 1. Introduction

It is known that many reactions in supernovae explosions proceed through bound states and low-energy resonance states. To calculate the rate of such reactions, one needs to find the asymptotic normalization coefficient (ANC) of the radial wave function for bound and resonance states, which can be used to calculate radiative capture cross sections at low energies. This process is one of the main sources of new element creation. An outstanding example of the important role of resonances is known for the unstable $^8$Be ground state. It presents a very narrow resonance with the pole at the center-of-mass system (cms) energy (see the review [1] and the references therein, $E_{\alpha\alpha}$ in eV)

$$E_{\alpha\alpha}= E_0 - i\, \Gamma/2 = 0.9184 \cdot 10^4 - i\, 2.8. \tag{1}$$

From the indeterminacy principle, the lifetime of the $^8$Be is $\tau = \hbar/\Gamma \approx 10^{-16}$ s. The lifetime of $^8$Be and the value $Q = E_0$ make the creation of $^{12}$C possible. Fred Hoyle predicted the existence of the resonance state of the $^{12}$C nucleus with an excitation energy of 7.68 MeV even before of 7.65 MeV was observed in experiments. Hoyle reasoned this from the natural occurrence of $^{12}$C in the universe (possibly 'the anthropic principle', see, for example, [2]).

Resonance states are described by the so-called Gamov wave functions that contain only the outgoing waves asymptotically, which exponentially increase due to the complex momenta $k_r$. Here $k_r = \sqrt{2\mu E_r}$, $k_r = k_0 − ik_i$, $E_r = E_0 − i\Gamma/2$ is a resonance energy, $\mu$ is a reduced mass of colliding part6icles. In the past (see [3] and references therein), the analytical continuation onto the unphysical energy sheet of the Lippmann–Schwinger as well as the momentum space Faddeev integral equations were used to find the resonance properties. The normalization formula for the bound state vertex function in the momentum space was generalized in [4] for the resonance and virtual states.

A potential S-matrix pole (PSMP) method was developed in Ref. [5] for obtaining the S-matrix pole parameters for bound, virtual and resonant states based on numerical solutions of the Schrodinger equation. In Ref. [6] a similar method was proposed for finding the resonance parameters of a nuclear system, obtained from the phase-shift analysis of elastic scattering data, by means of a pole representation of an S-matrix in complex momentum space. In Ref. [7], the S-matrix pole method (SMP) is suggested. An analytical approximation for the non-resonant part of the phase shift is given in the form of a series, which can be analytically continued to the point of an isolated resonance pole in the complex plane of the momentum $k$. In [7] the SMP and ERF methods results are compared for the $^5$He, $^5$Li and $^{16}$O resonance states. Different recent methods of the ANC estimation for bound and resonance states are discussed in Ref. [8].

In the present paper, the $\Delta$ method is applied, using the $\Delta_l$ function fittings for the experimental $p$-$^4$He scattering phase-shift data in the $P_{3/2}$ and $P_{1/2}$ resonance states. The important role of unstable nuclei in astrophysics is described in [9], where the $p$-$\alpha$ scattering phase shifts found by an accurate R-matrix analysis are presented. In the present paper, these phase-shifts are used as input data. The name $\Delta$ method is used here because the fit to experimental data energy dependence of the phase-shift is made for the $\Delta_l$ function. The Coulomb term is known in an analytical form and does not need fitting. In the literature the same name ($\Delta$ method) is used for the method proposed by Ramírez Suárez and Sparenberg in Ref. [10] where a bound state pole position is defined by the equation $\Delta_l$=0.

In Ref. [11] and in the present paper, the standard pole condition including the Coulomb term is used for a resonance (see Eqs. (16-17) below). In [11] the $\Delta$ method is successfully applied to the ANC calculation for low-energy resonance states of $^7$Be, $^8$Be, and $^{16}$O. This method avoids problems arising in the ERF and Padé approximation method when charges of colliding particles increase. These approaches do not work for large charges when the nuclear term of the ERF is too small, compared with the Coulomb term. The $\alpha^{12}$C system is a good example, when the nuclear part is on average three orders of magnitude smaller than the Coulomb part (see Ref. [11]).

To check the results in Ref. [11] and in the present work a simple formula, approximated ANC dependence on $E_0$ and $\Gamma$ for a narrow resonance, borrowed from Dolinsky and Mukhamedzhanov paper [12], which we call the DM method, is used. An alternative derivation of DM formula by the SMP method is given in Ref. [7].

An earlier study of $p$-$\alpha$ resonant scattering is given in Ref. [13]. The well-known expression (see, for example, Ref. [14]) for the ERF in the $P$ wave is used in Ref. [13] for finding energy poles of the $S$-matrix in the $N$-$\alpha$ scattering near the elastic-scattering threshold. However, the re-normalized nuclear vertex constants (NVC) and ANC of the Gamow wave functions are not considered in [13]. The first ERF method application for the ANC finding is made in Ref. [15]. In Ref. [16] the ERF method is generalized to resonance properties calculations. The N/D method is applied in Refs. [17], [18] for calculating the values of the parameters of the resonances in $N$-$\alpha$ scatterings.

A reasonable way to find when the ERF does not work is to compare the resonance energies and ANCs calculated by the ERF method with those obtained by the $\Delta$ method for the same input. It is necessary to consider systems lighter than oxygen. This begins in Ref. [11], where some results of the ERF method for the $\alpha$-$\alpha$ scattering are also compared with the $\Delta$ method results.

An analytical continuation from the physical region to the point situated in the fourth quadrant of the complex momentum (i. e. in the second energy unphysical sheet) is used to find the $\Delta_l$ function at a resonance pole. This means that the $\Delta_l$-function fitting for a concrete resonance should be especially good in the region around the real part $E_0$ of the resonance energy $E_r$=$E_0$-$i$ $\Gamma$/2.

The present article is organized as follows.

In Sec. 2 the main formulas of the $\Delta$ method for resonance parameters are given.

Sec. 3 gives the calculation results. The $\Delta_l$ function curves for the $P_{3/2}$ and $P_{1/2}$ states and Re $h(\eta)$ are shown in Fig. 1. The $\Delta_l(E)$ polynomial fittings up to $E^2$ and $E^4$ are compared. The exact

Re $h(\eta)$ is described nicely by the polynomial up to $E^2$ when the experimental region is the same as for the $P_{3/2}$ state.

The experimental $\Delta_l(E)$ energy dependence is more complex than the effective-range functions $K_l(E)$ behavior (see Fig. 2 in Ref. [7]). Polynomials up to the fourth degree of $E$ are needed for a good $\Delta_l(E)$ fitting. A table is given which includes the experimental and calculated resonant energy $E_0$ and width $\Gamma$ for the $\Delta$ method as well as the resulting absolute values of the re-normalized amplitude residue $W_l$, the nuclear vertex constant NVC and ANC. Larger differences between the results obtained by the ERF and $\Delta$ methods are seen in the resonance energy and width.

In Sec. 4 the main results of the paper are summarized. The results obtained show notable differences between the ERF and $\Delta$ method resonance parameters. The $\Delta$ method calculations are no more complicated than those of the ERF method, but give a more accurate scattering amplitude fitting and have no limits on collided particles charges. This means that the $\Delta$ method is preferable for the $^5$Li lowest two levels energies and ANCs estimations, although for this relatively light system the nuclear and Coulomb parts of $K_l(E)$ are of the same order of magnitude.

The unit system $\hbar = c = 1$ is used.

## 2. The $\Delta$ method for deducing the $^5$Li low-energy resonant state properties

The nucleus $^5$Li is interesting in that the ground and first excited states are resonance states, which can be treated as single-channel systems. The phase shift of the elastic $p$-$\alpha$ scattering for total angular momentum and parity $J^\pi = 3/2^-$ passes through $\pi/2$, and therefore leads to a narrow resonance. However, the phase shift of the elastic $p$-$\alpha$ scattering for $J^\pi = 1/2^-$ does not pass through $\pi/2$, so the corresponding resonance is quite wide. The $\Delta$ method is used to calculate the resonant energy, the residue $W_l$ of the re-normalized scattering amplitude, the nuclear vertex constant NVC and the ANC.

The re-normalized scattering amplitude, taking into account the Coulomb interaction, is derived in [19] to enable the analytic continuation of this amplitude to negative energies. The following notations are used below: $\eta = \xi/k$ is the Sommerfeld parameter, $\xi = Z_1 Z_2 \mu \alpha = 1/a_B$, $k = \sqrt{2\mu E_c}$ is the relative momentum; $\mu$ and $E_c$ are the reduced mass and the cms energy of the colliding nuclei with the charge numbers $Z_1$ an $Z_2$ respectively; $a_B$ is the Bohr radius and $\alpha$ is the fine-structure constant.

It is shown in [19] that in the physical energy sheet the ERF is a real analytic function with the possible exception of single poles. This means that the ERF can be described by the ERE or Padé approximations (see Refs. [20, 11]), whose coefficients can be found by fitting the experimental phase shifts. The same is valid for the $\Delta_l$ function because the nuclear part of the ERF, including the $\Delta_l$, is also a meromorphic function of energy in the physical area.

The partial amplitude of the nuclear scattering in the presence of the Coulomb interaction is (an index $cs$ means Coulomb+strong interactions)

$$f_l(k) = \exp(2i\sigma_l)[\exp(2i\delta_l^{(cs)}) - 1]/2ik, \tag{2}$$

where $\delta_l^{(cs)}$ is the Coulomb-nucleus phase shift,

$$\exp(2i\sigma_l) = \Gamma(l + 1 + i\eta)/\Gamma(l + 1 - i\eta). \tag{3}$$

As given in Ref. [19], the partial amplitude of the elastic scattering is re-normalized by multiplying it by the function

$$CF_l(k) = (l!)^2 \exp(\pi\eta)/[\Gamma(l + 1 + i\eta)]^2. \tag{4}$$

Thus, the expression for the re-normalized amplitude of the elastic scattering can be written as

$$\tilde{f}_l = 1/[k(\cot \delta_l^{(cs)} - i) \rho_l(k)], \tag{5}$$

where the function $\rho_l$ is defined by the equation

$$\rho_l(k) = 2\eta\, C_0^2(\eta)\, \prod_{n=1}^{l}(1 + \eta^2/n^2). \tag{6}$$

Here the following notation is used:

$$C_0^2(\eta) = \pi / [\exp(2\pi\eta) - 1]. \tag{7}$$

It is easy to recast (5) as

$$\tilde{f}_l = k^{2l} / [2\xi\, D_l(k^2)\, C_0^2(\eta)\, (\cot \delta_l^{(cs)} - i)]. \tag{8}$$

$$D_l(k^2) = \prod_{n=1}^{l}(k^2 + \xi^2/n^2), \quad D_0(k^2) = 1. \tag{9}$$

We define the $\Delta_l$ function as

$$\Delta_l = \pi \cot \delta_l^{(cs)} / [\exp(2\pi\eta) - 1] \tag{10}$$

in the positive energy semi-axis.

Writing the expression $\cot \delta_l$ in the non-physical energy region in Eq. (5) and elsewhere means its analytical continuation, since the phase shift is defined only in the positive energy region. Here and further down we omit the upper indexes in $\delta_l^{(cs)}$ for brevity.

The function $C_0^2(\eta)$, having the analytical form (7), does not need fitting. This function clearly depends on momentum $k$ through $\eta(k)$ which leads to the square root cut of the re-normalized amplitude in the complex energy plane. As mentioned above, the Coulomb part of the ERF leads to a much smoother energy dependence $K_l(k^2)$ compared with $\Delta_l(k)$. The standard expression for $K_l(k^2)$ [14] is given by the following equations:

$$K_l(k^2) = 2\xi\, D_l(k_r^2)[C_0^2(\eta)\, (\cot \delta_l - i) + h(\eta)], \tag{11}$$

$$h(\eta) = \Psi(\eta) + (2i\eta)^{-1} - \ln(i\eta). \tag{12}$$

In [11] the ERF fitting for the $N$-$\alpha$ scattering is given by the standard equation

$$K_l(k^2) = -1/a_l + (r_l/2)k^2 - P_l r_l^3 k^4 \tag{13}$$

which is equivalent to the following equation, where $E_c$ is the energy of colliding nuclei in the cms system:

$$K_l(E_c) = b_0 + b_1 E_c + b_2 E_c^2. \tag{14}$$

The experimental points $\Delta_J(E_c)$ have a more sophisticated distribution compared with the smooth energy dependence of $K_l(E_c)$. $\Delta_J(E_c)$ increases when energy decreases. In the region $E_c > 6.5$ MeV, the decrease in functions $\Delta_J(E_c)$ for $J=3/2$ and $1/2$ is replaced by an increase. Due to this, good $\Delta_J(E_c)$ fittings are obtained using polynomials in powers of $E_c$ up to $E_c^4$ including five fitting parameters instead of three as in (14):

$$\Delta_J(E_c) = a_{0J} + a_{1J} E_c + a_{2J} E_c^2 + a_{3J} E_c^3 + a_{4J} E_c^4. \tag{15}$$

In (15) and below $J$ is used instead of $l$ to indicate the state because of the fixed $l=1$ and the different $J = 1/2, 3/2$. To find the resonance energy position the standard equation is used:

$$\cot \delta_l - i = 0. \tag{16}$$

With the fitted parameters of $\Delta_J(E_c)$ in (15), the resonance complex energy can be found from the equation

$$\Delta_J(E_c) - iC_0^2(\eta) = 0 \tag{17}$$

which is equivalent to Eq. (16), and then the listed values of constants calculated. Our new results are presented in Table 1 in the $\Delta$ lines.

The values of the residue of $\tilde{f}_l(k)$ at the resonance energy $E_r$ for the $\Delta_l$ fitting can be written as

$$W_l(k_r) = k_r^{2l} \Big/ \Big\{ 2\xi \, D_l(k_r^2) \lim_{k \to k_r} \Big( \frac{d}{dk} [\Delta_l(k^2) - i\, C_0^2(\eta)] \Big) \Big\}. \tag{18}$$

(see [11]). According to the known relations between the NVC ($\tilde{G}_l$), ANC ($C_l$) and the residue $W_l$ one can write

$$\tilde{G}_l^2 = -(2\pi k_r/\mu^2)\, W_l, \tag{19}$$

$$C_l = (i^{-l}\mu/\sqrt{\pi})\, [\Gamma(l+1+i\eta_r)/l!] \exp(-\pi\eta_r/2)\, \tilde{G}_l \tag{20}$$

where $\eta_r = \xi/k_r$. The simple DM relation for the ANC derived in Ref. [12]

$$|C_l^a| = \sqrt{\mu\Gamma/k_0} \tag{21}$$

is used to check the calculations for the narrow resonance in the ground state $J^\pi = 3/2^-$.

## 3. The calculation results

The function $\Delta_J = \Delta(J)$ fitting curves, which are obtained using polynomial in powers of $E$ up to the $E^4$ (15) (solid lines which run almost through experimental points) and that up to $E^2$ (dashed lines), are shown in Fig. 1.

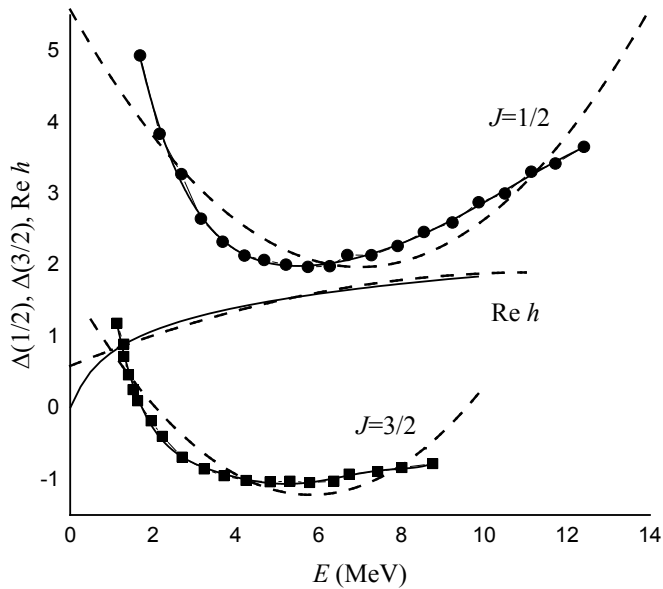

Fig. 1. Experimental Δ(J) points for $J^\pi$ = 3/2$^-$ (dots) and for $J^\pi$ = 1/2$^-$ (squares) vs the cms energy $E=E_c$. Polynomials up to $E^4$ (solid lines) and up to $E^2$ (dashed curves) for $J^\pi$ = 3/2$^-$ and $J^\pi$ = 1/2$^-$ are used for fitting the experimental Δ(J) behavior. The middle lines without symbols are the Coulomb Re $h$ function (solid) and its fitting by the polynomial up to $E^2$ (dashed line).

The dashed lines for Δ functions relate directly to the EFR method, since the same set of experimental data is used for both methods. They deviate notably from the experimental points, while the function Re $h$ is well-described by a polynomial of type (13) in the selected energy region. The solid and dashed Δ curves begin to diverge to the left of the last points where experimental data uncertainties increase due to the Coulomb barrier. Therefore, the corresponding points are not included in the fittings. The selected energy intervals include $E_0$ for both resonances. A similar divergence between solid and dashed curves takes place in the region to the right of these intervals. However, this region does not significantly affect the resonance parameters because it is far from the resonance energies. As mentioned above, the most important region for an analytical continuation of Δ functions and data fitting is the area around the resonance position $E_0$. The resulting $E_0$ values are at close proximity to the intersections of the solid and dashed curves for both states (see Table 1). The crossing point is 1.46 MeV for J=3/2 ($E_0$= 1.39 for Δ method, $E_0$= 1.48 for ERF) and 2.40 for J=1/2 ($E_0$= 2.61 for Δ method, $E_0$= 2.41 for ERF). This leads to a better agreement between the resonance parameters found for the ERF and Δ methods. Nevertheless, differences still exist which are stronger for $E_0$ and Γ compared with ANC.

Table I.
The p-α scattering resonances. Methods: ERF [7] and Δ (present paper)]; $J^\pi$; resonance energy $E_0$ in the center-of-mass system (in MeV); corresponding width Γ; the residue $|W_l|$; NVC $|\tilde{G}_l^2|$; and ANCs denoted as $|C_l|$ and $|C_l^a|$ (for narrow resonances [12]). The fitting models are described in the text. The phase shifts are borrowed from Ref. [9].

| Method | $J^\pi$ | $E_0$ [MeV] | Γ [MeV] | $|W_l|$ | $|\tilde{G}_l^2|$ [fm] | $|C_l|$ [fm$^{-1/2}$] | $|C_l^a|$ [fm$^{-1/2}$] |
|---|---|---|---|---|---|---|---|
| ERF | 3/2$^-$ | 1.481 | 1.041 | 0.295 | 0.0314 | 0.260 | 0.288 |
| Δ | | 1.390 | 1.301 | 0.393 | 0.0416 | 0.297 | 0.325 |
| ERF | 1/2$^-$ | 2.213 | 4.640 | 0.305 | 0.0459 | 0.357 | 0.526 |
| Δ | | 2.611 | 4.534 | 0.223 | 0.0355 | 0.314 | 0.505 |

The following conclusions can be drawn from Table I.
The numerical differences for the resonant parameters in the ERF and Δ methods are as follows: 6.5% in the J=3/2 state and 15% in the J=1/2 state for $E_0$; 20% in the J=3/2 state and 2.4% in the J=1/2 state for Γ; 12.5% in the J=3/2 state and 14% in the J=1/2 state for the ANC $|C_l|$. The differences between the ANCs ($|C_l|$ and $|C_l^a|$) in the J=3/2 state are 11% for the ERF method and 9% for the Δ method. The same difference in the J=1/2 state are 47% for the ERF method and 61% for the Δ method. The related differences are much smaller for the narrow resonance in the J=3/2 state. The resonance energy (especially Γ for the narrow level) is more sensitive to the method than the ANC, which agrees with Eq. (21). In Ref. [11] it is noted that a larger sensitivity of Γ compared with $E_0$ is a consequence of the fact that the ratio $\Gamma/\sqrt{E_0}$ appears in (21). The related uncertainties of $E_0$ and Γ are given in [11].

## 4. Conclusion

In the present paper, the properties of the ground 3/2$^-$ and first excited 1/2$^-$ resonances in p-$^4$He scattering are studied, using the recent phase-shift from Ref. [9] as input data. The resonance parameters calculated using the ERF and Δ methods are compared in Table 1. The differences between resonance energies are significant: up to 15% for $E_0$ and 20% for Γ. These values exceed uncertainties in the phase shifts.

In our opinion, the Δ method of the ANC estimation is preferable to the ERF method whose results are still important as a first reasonable estimation of resonance parameters. The Δ method is no more difficult to calculate than the EFR method, it has no limits on the charges of collided particles, and guarantees a correct description of the Δ function. Refining the experimental $p$-$^4$He phase shifts to receive more accurate resonance parameters is needed. The results of the present paper show that a recalculation of the ERF method results using the Δ method is useful for systems lighter than $^{16}$O. The resonance energy and ANC results can be applied in nuclear astrophysics for radiative capture cross section calculations, and in the direct nuclear reactions theory for the estimation of the vertex part in the Feynman diagram.

**Declaration of competing interest**

The author declares that he has no known competing financial interests or personal relationships that could have appeared to influence the work reported in this paper.

**Acknowledgements**

This work was partially supported by the Russian Science Foundation (Grant No. 19-02-00014). The author is grateful to H. M. Jones for editing the English.